\documentclass[letterpaper]{article}

\usepackage{mathtools}
\usepackage{a4wide}
\usepackage{standalone}
\usepackage[justification=centering]{subcaption}
\usepackage{setspace}
\usepackage{tikz}
\usetikzlibrary{calc}
\usetikzlibrary{matrix}
\usetikzlibrary{arrows}
\usetikzlibrary{positioning}

\title{An informational study of the evolution of codes and of emerging concepts in populations of agents}
\author{Andr\'{e}s C. Burgos$^1$ \and Daniel Polani$^2$ \\
\mbox{}\\
$^{1,2}$Adaptive Systems Research Group, University of Hertfordshire, Hatfield, UK \\
$^1$Email address: a.c.burgos@herts.ac.uk, Tel: +44 1707 28 4490}
\date{}

\begin{document}
\maketitle

\begin{abstract}
We consider the problem of the evolution of a code within a structured population of agents. The agents try to maximise their information
about their environment by acquiring information from the outputs of other agents in the population. A naive use of information-theoretic
methods would assume that every agent knows how to ``interpret'' the information offered by other agents. However, this assumes that one
``knows'' which other agents one observes, and thus which code they use. In our model, however, we wish to preclude that: it is not clear
which other agents an agent is observing, and the resulting usable information is therefore influenced by the universality of the code used
and by which agents an agent is ``listening'' to. We further investigate whether an agent who does not directly perceive the environment can
distinguish states by observing other agents' outputs. For this purpose, we consider a population of different types of agents ``talking''
about different concepts, and try to extract new ones by considering their outputs only.
\\ \\
\smallskip
\noindent \textbf{Keywords:} information theory, code evolution, semantics, emerging concepts
\end{abstract}

\newpage
\section{Introduction}
\label{sec:introduction}

If we consider organisms capable of processing information, then we can argue that they must be able to internally assign meaning to the
symbols they perceive in a code-based manner \cite{Gorlich2011}. For instance, bacteria perceives chemical molecules in their environment
and interprets them in order to better estimate environmental conditions and (stochastically) decide their phenotype \cite{Platt2010, Balazsi2011,
Perkins2009, Schuster2013}. Plants detect airborne signals released by other plants, being able to interpret them as attacks of pathogens or herbivores
\cite{Heil2010, Shah2009}. Therefore, a correspondence between environmental conditions and chemical molecules must be established. It is in this way
that Barbieri characterises codes, and he proposes three fundamental characteristics for them: they connect two independent worlds; they add
meaning to information; and they are community rules \cite{Barbieri2003}.

Codes connect two independent worlds by establishing a correspondence or mapping between them. These worlds are independent and thus there
are no material constraints for establishing arbitrary mappings. The meaning of information comes exclusively from the mapping: symbols by
themselves are meaningless. Finally, the third property requires that the correspondence between the two worlds constitutes an integrated system.

For instance, human languages establish a correspondence between words and objects \cite{Barbieri2003}; in bacteria it is between chemical
molecules and environmental and social conditions \cite{Waters2005, West2006}. Words (or chemical molecules) by themselves do not have any meaning, and
each individual of a population can define, arbitrarily to some extent, their own set with its mapping. However, populations of individuals sharing
the same code are ubiquitous in nature. How is it that codes come to be shared by many individuals when their constitution involve arbitrary choices
for each individual? This question is what we are investigating in the present paper.

For this work, we assume a simple scenario where organisms live in a fluctuating environment. If they can perfectly predict the future
environmental conditions, they can prepare themselves by adopting a proper phenotype, and, therefore, survive. However, when uncertainty about
the environment remains, organisms will follow a bet-hedging strategy \cite{Slatkin1974, Seger1987}, where they try to maximise their long-term
growth rate by adopting the phenotype that matches the environment in proportions based on the information they have about it. For example, seeds
of annual plants germinate stochastically in different periods in relation to the probability of rainfalls, and their chances of survival are
maximised when they match this probability \cite{Cohen1966}.

The relation between information and long-term growth rate can be expressed elegantly in information theoretic terms, where an increase in the
environmental information of an organism is translated into an increase in its long-term growth rate \cite{Shannon1948, Kelly1956, Kussell2005,
DonaldsonMatasci2010, Rivoire2011}. Such models achieve the maximisation of the long-term growth rate by maximising an organism's information about
the environment. If we assume this behaviour in organisms, then those obtaining additional environmental information (other than that from their sensors,
which we assume it does not completely eliminate environmental uncertainty) from other individuals will have an advantage over those that do not,
since they would be able to better predict the future conditions. However, for individuals to be able to communicate with each other, they must be
able to translate symbols into environmental conditions, where the output of these symbols results from an individual's code. We consider the code
of an individual as a stochastic mapping from its sensors states to a set of outputs.

For this study, we consider outputs (or messages) of individuals (or agents) as conventional signs. In semiotics, the science of all processes in which
signs are originated, stored, communicated, and being effective \cite{Gorlich2011}, two types of signs are traditionally recognised: \emph{conventional
signs} and \emph{natural signs} \cite{Deely2006}. In conventional signs there is no physical constraint on the possible mappings, they are established by
conventions. Although in physical systems there can be limitations to the possible mappings that can be implemented, in this work we assume complete
freedom of choice. On the other hand, in natural signs, there is always a physical link between the signifier and signified, such as smoke as a sign
of fire, odours as signs of food, etc. \cite{Barbieri2008}.

In this work, we are not interested in the particular detailed mechanisms by which an agent implements its code, nor how the agent decodes the outputs
of other agents. Instead, we focus on the theoretical limits on the amount of environmental information an agent can possibly acquire resulting from
different scenarios of population structure and codes distribution. 
The natural framework to analyse such quantities is information theory \cite{Shannon1948}. However, it does not take semantic aspects into account,
it only deals with frequencies of symbols instead of what they symbolise. Codes, on the other hand, add meaning to information, which makes the
integration of sciences such as semiotics with information theory non-trivial \cite{Favareau2007, Battail2009}. In the following section, we present
an information-theoretic model which incorporates the necessity of conventions by dropping from the model the usual implicit assumption of knowing
the identity of the communicating units.

\section{Model}

\begin{figure}[ht]
\centering
  \begin{tikzpicture}
    [->,>=stealth',shorten >=2pt,auto,node distance=1.5cm,
    thick,main node/.style={font=\sffamily\normalsize\bfseries}]

    \node[main node] (2) [] {$\mu$};
    \node[main node] (5) [below of=2] {$Y_{\theta_2}$};
    \node[main node] (3) [left of=5] {$Y_{\theta_1}$};
    \node[main node] (4) [right of=5] {$Y_{\theta_3}$};
    \node[main node] (6) [below of=5] {$X_{\theta_2}$};
    \node[main node] (7) [below of=3] {$X_{\theta_1}$};
    \node[main node] (8) [below of=4] {$X_{\theta_3}$};

    \path[every node/.style={font=\sffamily\small}]
      (2) edge node {} (3)
      (2) edge node {} (4)
      (2) edge node {} (5)
      (3) edge node {} (7)
      (4) edge node {} (8)
      (5) edge node {} (6)
      ;
  \end{tikzpicture}
  \caption{\doublespacing \small{Bayesian network representing the relantionship between the sensor and output variables of three agents.}}
  \label{fig:naivecopymodel}
\end{figure}
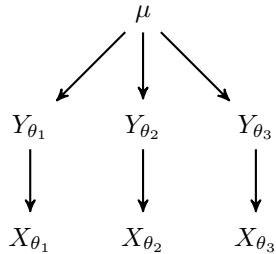

To introduce the model in a progressive manner, let us first consider three agents, $\theta_1$, $\theta_2$ and $\theta_3$. Each of these
agents depend on the same environmental conditions for survival, which are modelled by a random variable $\mu$. Agents acquire information
about the environment through their sensors, which are modelled by random variables $Y_{\theta_1}$, $Y_{\theta_2}$ and $Y_{\theta_3}$, all
three conditioned on $\mu$, for agents $\theta_1$, $\theta_2$ and $\theta_3$, respectively. We assume each agent acquires the same amount
and aspects of environmental information from $\mu$, \emph{i.e.} $p(Y_{\theta_1}|\mu) = p(Y_{\theta_2}|\mu) = p(Y_{\theta_3}|\mu)$. Let us
further assume that the information each agent acquires about the environment does not eliminate its uncertainty, \emph{i.e.}
$H(\mu|Y_{\theta_i}) > 0$ for $1 \le i \le 3$. The code of an agent is a stochastic mapping from its sensor states into a set of outputs,
and is represented by the conditional probabilities $p(X_{\theta_1}|Y_{\theta_1})$, $p(X_{\theta_2}|Y_{\theta_2})$ and
$p(X_{\theta_3}|Y_{\theta_3})$ for agents $\theta_1$, $\theta_2$ and $\theta_3$, respectively (see Fig. \ref{fig:naivecopymodel}). 

Let us assume that agent $\theta_1$ perceives only the outputs of agents $\theta_2$ and $\theta_3$. One possible way of computing the
information about the environment agent $\theta_1$ has is to consider the mutual information between $\mu$ and the joint distribution
of the sensor of $\theta_1$ and the outputs of $\theta_2$ and $\theta_3$: $I(\mu;Y_{\theta_1},X_{\theta_2},X_{\theta_3})$. However,
by writing down this quantity, we are implicitly assuming that agent $\theta_1$ ``knows'' which output corresponds to $\theta_2$ and which
output corresponds to $\theta_3$. Therefore, in this consideration, an agent can theoretically do the translations of the outputs according
to some internal model of other agents and infer the mentioned amount of information about its environment.

\subsection{Indistinguishable sources of messages}
\label{sec:indsources}

For this study, on the contrary, we consider an agent observing other agents' messages, but under the assumption that the originator of a
message cannot be identified. In this way, the total amount of information an agent can infer from the outputs of other agents will depend
on to which extent it either can identify who the other agents are or can rely on them using a coding scheme that does not depend too much
on their particular identity. For instance, if agents $\theta_2$ and $\theta_3$ both agree on the output for each of the environmental
conditions, then agent $\theta_1$ should be able to infer more environmental information than if they disagree on the output for each of the
environmental conditions, given that agent $\theta_1$ does not know which of the agents it is observing.

\begin{figure}[ht]
\centering
  \begin{tikzpicture}
    [->,>=stealth',shorten >=2pt,auto,node distance=1.8cm,
    thick,main node/.style={font=\sffamily\normalsize\bfseries}]

    \node[main node] (2) [] {$\mu$};
    \node[main node] (3) [below left of=2] {$Y_{\theta_1}$};
    \node[main node] (4) [below right of=2] {$Y_{\Theta^\prime}$};
    \node[main node] (7) [below of=3] {$X_{\theta_1}$};
    \node[main node] (8) [below of=4] {$X_{\Theta^\prime}$};
    \node[main node] (6) [right of=8] {$\Theta^\prime$};

    \path[every node/.style={font=\sffamily\small}]
      (2) edge node {} (3)
      (2) edge node {} (4)
      (6) edge node {} (4)
      (4) edge node {} (8)
      (6) edge node {} (8)
      (3) edge node {} (7)
      ;
  \end{tikzpicture}
  \caption{\doublespacing\small{Bayesian network representing the relationships as described above (see text).}}
  \label{fig:partialcopymodel}
\end{figure}
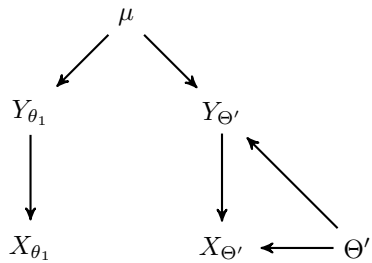

To model this idea, let us assume a random variable $\Theta^\prime$ denoting the selected agent. This agent depends on the same environmental
conditions for survival as $\theta_1$, which are modelled, as above, by a random variable $\mu$. Agents acquire information about the environment
through their sensors, which are modelled by a random variable $Y_{\Theta^\prime}$ conditioned on the index variable denoting the agent under consideration,
$\Theta^\prime$, and $\mu$. The amount of acquired sensory information of a specific agent $\theta^\prime$ about $\mu$ is given by
$I(\mu;Y_{\theta^\prime})$. As above, the code of an agent is a stochastic mapping from its sensor states into a set of messages, and is represented
by the conditional probability $p(X_{\theta^\prime}|Y_{\theta^\prime})$ for an agent $\theta^\prime$ (see Fig. \ref{fig:partialcopymodel}).

However, now we want to model the fact that we do not know which agent is observed. In the case with maximum uncertainty, $\Theta$ is uniformly
distributed, and then this parametrisation of the codes considers the outputs of all agents in $\Theta^\prime$ altogether, such that if we are not
observing $\Theta^\prime$, we cannot identify whose agent's output we are observing. In Eq. \ref{eq:condprobxtheta} and Eq. \ref{eq:condprobxthetaopp}
we show two examples of codes for agents $\theta_2$ and $\theta_3$, while their sensor states are define by the Eq. \ref{eq:condprobytheta} (Eq.
\ref{eq:condprobytheta1} defines the sensors states of agent $\theta_1$). We compute how much information about the environment there is when the
outputs of both agents ($\theta_2$ and $\theta_3$) are considered together by agent $\theta_1$.

\begin{figure}[htp]
	\centering
	\begin{minipage}[b]{.49\textwidth}
		\begin{equation}
			Pr(Y_{\theta_1}|\mu) \coloneqq 
			\bordermatrix{~ & y_1 & y_2 \cr
			              \mu_1 & 1 - \epsilon & \epsilon \cr
			              \mu_2 & \epsilon & 1 - \epsilon \cr}
			\label{eq:condprobytheta1}
		\end{equation}
	\end{minipage}\hfill
	\begin{minipage}[b]{.49\textwidth}
		\begin{equation}
			Pr(Y_{\Theta^\prime}|\mu, \Theta^\prime) \coloneqq 
			\bordermatrix{~ & y_1 & y_2 \cr
			              \theta_2,~\mu_1 & 1 - \epsilon & \epsilon \cr
			              \theta_2,~\mu_2 & \epsilon & 1 - \epsilon \cr
			              \theta_3,~\mu_1 & 1 - \epsilon & \epsilon \cr
			              \theta_3,~\mu_2 & \epsilon & 1 - \epsilon \cr}
			\label{eq:condprobytheta}
		\end{equation}
	\end{minipage}
\end{figure}

\begin{figure}[htp]
	\centering
	\begin{minipage}[b]{.49\textwidth}
		\begin{equation}
			Pr(X_{\Theta^\prime}|Y_{\Theta^\prime}, \Theta^\prime) \coloneqq 
			\bordermatrix{~ & x_1 & x_2 \cr
			             \theta_2,~y_1 & 1 & 0 \cr
			             \theta_2,~y_2 & 0 & 1 \cr
			             \theta_3,~y_1 & 1 & 0 \cr
			             \theta_3,~y_2 & 0 & 1 \cr}
			\label{eq:condprobxtheta}
		\end{equation}
	\end{minipage}\hfill
	\begin{minipage}[b]{.49\textwidth}
		\begin{equation}
			Pr(X_{\Theta^\prime}|Y_{\Theta^\prime}, \Theta^\prime) \coloneqq 
			\bordermatrix{~ & x_1 & x_2 \cr
		              \theta_2,~y_1 & 1 & 0 \cr
		              \theta_2,~y_2 & 0 & 1 \cr
		              \theta_3,~y_1 & 0 & 1 \cr
		              \theta_3,~y_2 & 1 & 0 \cr}
			\label{eq:condprobxthetaopp}
		\end{equation}
	\end{minipage}
\end{figure}

If we assume $p(\theta_2) = p(\theta_3) = 1/2$, and $p(\mu_1) = p(\mu_2) = 1/2$ and $\epsilon = 0.01$, then if we consider the codes shown in Eq.
\ref{eq:condprobxtheta}, we have that $I(\mu;Y_{\theta_1},X_{\Theta^\prime}) = 0.97872$ bits, where $\Theta^\prime$ consists of agents $\theta_2$
and $\theta_3$. However, had $\theta_2$ and $\theta_3$ ``opposite'' codes as shown in Eq. \ref{eq:condprobxthetaopp}, then
$I(\mu;Y_{\theta_1},X_{\Theta^\prime}) = 0.9192$ bits, which is exactly $I(\mu;Y_{\theta_1})$, that is, $I(\mu;X_{\Theta^\prime}|Y_{\theta_1}) = 0$
bits (agent $\theta_1$ cannot acquire any side information from the outputs of agents $\theta_2$ and $\theta_3$). We should note here that the
sensor states $y_1$ and $y_2$ of agents $\theta_2$ and $\theta_3$ in the conditional probability shown in Eq. \ref{eq:condprobytheta1} and
\ref{eq:condprobytheta} refer almost deterministically to the same environmental condition, and therefore the loss of side information is
thus entirely due to the incompatible codes. The conditional probabilities of sensor states given the environmental conditions further defined
throughout the paper are also assumed to be almost deterministically.

\subsection{Environmental information of a population}

The model shown in Fig. \ref{fig:partialcopymodel} considers the environmental information of agent $\theta_1$, ignoring its own output
$X_{\theta_1}$. Nevertheless, agents ignoring their outputs is contrary to our assumption over the incapability of agents to identify the
sources of the outputs. On the other hand, we are assuming a specific type of communication, one which could be classified as persistent
within the different classifications of stigmergy (\cite{Wilson2000, Theraulaz1999, Parunak2006}, see \cite{Heylighen2011} for a summary).
To incorporate this option in the model shown in Fig. \ref{fig:partialcopymodel}, we could consider the state space of $\Theta^\prime$ as
the set $\{ \theta_1, \theta_2, \theta_3 \}$. Then, to express not only the environmental information of agent $\theta_1$, but the average
environmental information of the whole population, we can parametrise the agent by a random variable $\Theta$ (defined over the same state
space, representing the same set of agents as $\Theta^\prime$), such that  $p(Y_\Theta|\mu, \Theta) = p(Y_{\Theta^\prime}|\mu, \Theta^\prime)$
(\emph{i.e.}, $Y_{\Theta^\prime}$ is \emph{i.i.d.} to $Y_\Theta$, and vice versa).

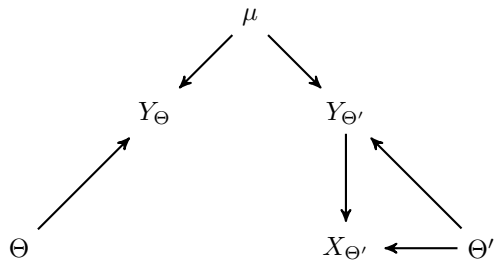
\begin{figure}[ht]
\centering
  \begin{tikzpicture}
    [->,>=stealth',shorten >=2pt,auto,node distance=1.8cm,
    thick,main node/.style={font=\sffamily\normalsize\bfseries}]

    \node[main node] (2) [] {$\mu$};
    \node[main node] (3) [below left of=2] {$Y_{\Theta}$};
    \node[main node] (4) [below right of=2] {$Y_{\Theta^\prime}$};
    \node[main node] (8) [below of=4] {$X_{\Theta^\prime}$};
    \node[main node] (6) [right of=8] {$\Theta^\prime$};
    \node[main node] (5) [below of=3] {};
    \node[main node] (7) [left of=5] {$\Theta$};

    \path[every node/.style={font=\sffamily\small}]
      (2) edge node {} (3)
      (2) edge node {} (4)
      (6) edge node {} (4)
      (4) edge node {} (8)
      (6) edge node {} (8)
      (7) edge node {} (3)
      ;
  \end{tikzpicture}
  \caption{\doublespacing\small{Bayesian network representing the sensor variables of a set of agents indexed by the random variable $\Theta$, and the sensor
			and output variables of a copy of the set of agents indexed by $\Theta$ named $\Theta^\prime$.}}
  \label{fig:partialcopymodel2}
\end{figure}

In this way, the average environmental information of a population of the agents selected by $\Theta$ is given by $I(\mu;Y_\Theta,X_{\Theta^\prime})$
(see Fig. \ref{fig:partialcopymodel2}). This measure can be consider as the objective function to maximise in our model. However, we
would be making two important assumptions: first, this objective function assumes agents have access to the environmental conditions
$\mu$, which they indirectly do but only through their sensors; and second, every agent would perceive the output of every other agent,
including itself. In this work, instead, we propose that agents follow a behaviour such that it maximises the similarity of their outputs
(via their codes) with those of which the agent perceives. A consequence of this behaviour is that the average information about $\mu$ is
also maximised. In addition, we will introduce a potentially flexible ``population structure'', so that we can specify which agents interact
with which.

\subsection{Code similarity}

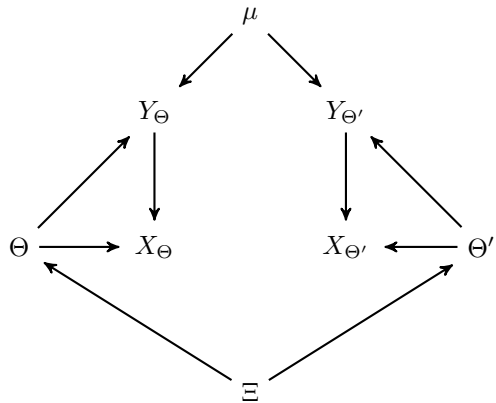
\begin{figure}[ht]
\centering
  \begin{tikzpicture}
    [->,>=stealth',shorten >=2pt,auto,node distance=1.8cm,
    thick,main node/.style={font=\sffamily\normalsize\bfseries}]

    \node[main node] (2) [] {$\mu$};
    \node[main node] (3) [below left of=2] {$Y_{\Theta}$};
    \node[main node] (4) [below right of=2] {$Y_{\Theta^\prime}$};
    \node[main node] (7) [below of=3] {$X_{\Theta}$};
    \node[main node] (8) [below of=4] {$X_{\Theta^\prime}$};
    \node[main node] (5) [left of=7] {$\Theta$};
    \node[main node] (6) [right of=8] {$\Theta^\prime$};
    \node[main node] (1) [below of=2, node distance=5cm] {$\Xi$};

    \path[every node/.style={font=\sffamily\small}]
      (1) edge node {} (5)
      (1) edge node {} (6)
      (2) edge node {} (3)
      (2) edge node {} (4)
      (5) edge node {} (3)
      (6) edge node {} (4)
      (3) edge node {} (7)
      (4) edge node {} (8)
      (5) edge node {} (7)
      (6) edge node {} (8)
      ;
  \end{tikzpicture}
  \caption{\doublespacing\small{Bayesian network representing the relantionship of the variables in the model of code evolution. $Y_{\Theta^\prime}$ is an \emph{i.i.d}
		copy of $Y_\Theta$ and $X_{\Theta^\prime}$ is an \emph{i.i.d.} copy of $X_\Theta$. $\Theta^\prime$ covers the same set of agents as
		$\Theta$, but its probability distribution is not necessary the same.}}
  \label{fig:encmodel}
\end{figure}

First, we introduce a copy of the codes of the agents, such that when we instantiate the variables $X_\Theta$ and $X_{\Theta^\prime}$,
the probabilities are the same. The structure of the population is then given by $p(\Theta,\Theta^\prime) = p(\Theta) p(\Theta^\prime)$.
However, the conditional independence of $\Theta$ and $\Theta^\prime$ restricts significantly the diversity of the structures that can be
represented. In such cases, the agents selected by $\Theta$ perceive the outputs of \emph{all} the agents selected by $\Theta^\prime$ and
vice versa. In order to model a general interaction structure between agents, we consider $p(\Theta,\Theta^\prime)$ not independent, as
shown in the Bayesian network in Fig. \ref{fig:encmodel}, where we introduce a helper variable $\Xi$. This allows different agents selected
by $\Theta$ to perceive outputs from exclusive agents selected by $\Theta^\prime$.

We define the objective function as $I(X_\Theta;X_{\Theta^\prime})$, that is the average code similarity of a population of agents according
to the population structure $p(\Theta, \Theta^\prime)$. For instance, if the interaction probability of two agents is zero, then the similarity
of the codes of these two agents is irrelevant for the objective function. On the other hand, they interact with probability bigger than zero
($p(\theta, \theta^\prime) > 0$, for some agents $\theta$ and $\theta^\prime$), then how similar their codes are will influence
$I(X_\Theta;X_{\Theta^\prime})$.

If we consider our system as a process in time, then at each time-step two agents are chosen according to $p(\Theta, \Theta^\prime)$. Agent
$\Theta$ reads the output of agent $\Theta^\prime$ (generated via its code, which is \emph{i.i.d} over time), and let us assume that it stores
the pair $(Y_\Theta,X_{\Theta^\prime})$, \emph{i.e.} its current sensor state together with the perceived output. If this is repeated a large
number of times, then the total amount of environmental information that can be inferred from the collected statistics by the population is
bounded by $I(\mu;Y_\Theta,X_{\Theta^\prime})$. This is the theoretical limit to which we refer in the introduction, and for this study we
are not interested in how the inference is computed. However, we implicitly assume that agents decode the perceived outputs according to their
codes.

\subsection{Distance between two codes}

In order to visualise the evolution of codes, we define the distance between the codes of two agents $\theta_i$ and $\theta_j$ as the square root of the
\emph{Jensen-Shannon divergence} \cite{Wong1985, Lin1991} between them. This measure has the property that $0 \leq JSD(\theta_i, \theta_j) \leq 1$
when $\log_2$ is used, and the square root yields a metric.
Let us note that this distance requires the sensor states $Y$ to be named identically (for the corresponding states of $\mu$) among agents
in order to be meaningful. As we stated above, this is (closely) the case in all our experiments. This requirement over the sensor states
discards the possibility of using other measures such as mutual information.

\begin{align}
dist(\theta_i,\theta_j) &= \sqrt{JSD\big(p(X_{\theta_i}|Y_{\theta_i})||p(X_{\theta_j}|Y_{\theta_j})\big)} \\ 
						&= \sqrt{\frac{1}{2} D\big(p(X_{\theta_i}|Y_{\theta_i})||p(X_{\theta_k}|Y_{\theta_k})\big)  
						+ \frac{1}{2} D\big(p(X_{\theta_j}|Y_{\theta_j})||p(X_{\theta_k}|Y_{\theta_k})\big)} \nonumber
\end{align}

where $p(X_{\theta_k}|Y_{\theta_k}) = \frac{1}{2} \big(p(X_{\theta_i}|Y_{\theta_i})+p(X_{\theta_j}|Y_{\theta_j})\big)$. 

\section{Methods}

To illustrate the behaviour of our model, we consider four different scenarios, which are described in Sec. \ref{sec:results}. The common
parameters for the first two experiments are the following: the population consists of $25$ agents; the amount and quality of the acquired
sensory information is the same for every agent, that is $p(Y_\Theta|\mu) = p(Y_{\Theta^\prime}|\mu)$. For the third scenario, the only
difference is that we consider only $15$ agents, since the dimensions to consider with a flexible structure grows quadratically with the
number of agents.

The optimisation algorithm used in the following experiments is CMA-ES (Covariance Matrix Adaptation Evolution Strategy), which is a
stochastic derivative-free method for non-linear optimisation problems \cite{Hansen2001}. We utilised the implementation provided by
the Shark library v3.0.0 \cite{Igel2008} with its default parameters, which implements the CMA-ES algorithm described in \cite{Hansen2004}.
The evolutionary algorithm used for optimisation does not intend to represent the actual evolution of the codes. Instead, we are interested
in the solutions of this optimisation process, which are representative of the possible outcomes of evolution.

To visualise the evolution of the codes of the agents, we use the method of multidimensional scaling provided by R version 2.14.1
(2011-12-22). This method takes as input the distance matrix between codes, and plots them in a two-dimensional space preserving the distances
as well as possible. To visualise, not only the distances between the resulting codes, but also how they relate to the distances between initial
codes, we provide a distance matrix of both initial and resulting codes. The initial codes are randomly set by the evolutionary algorithm.

\section{Results}
\label{sec:results}

In this section, we analyse the outcome of the four different scenarios where code similarity is maximised. While the outcomes are particular
for one simulation, they are illustrative of the richness that the model is able to capture, which is described for each scenario. The
outcomes are typical solutions, and we cannot perform statistics over simulations since the many solutions are qualitatively different.
However, the outcome of each scenario is presented together with a description of alternative outcomes, giving indicators of achievement
of local/global optimum.

\subsection{Well-mixed population}
\label{sec:allvsall}

\begin{figure}[htp]
\centering
	\includegraphics[scale=0.4]{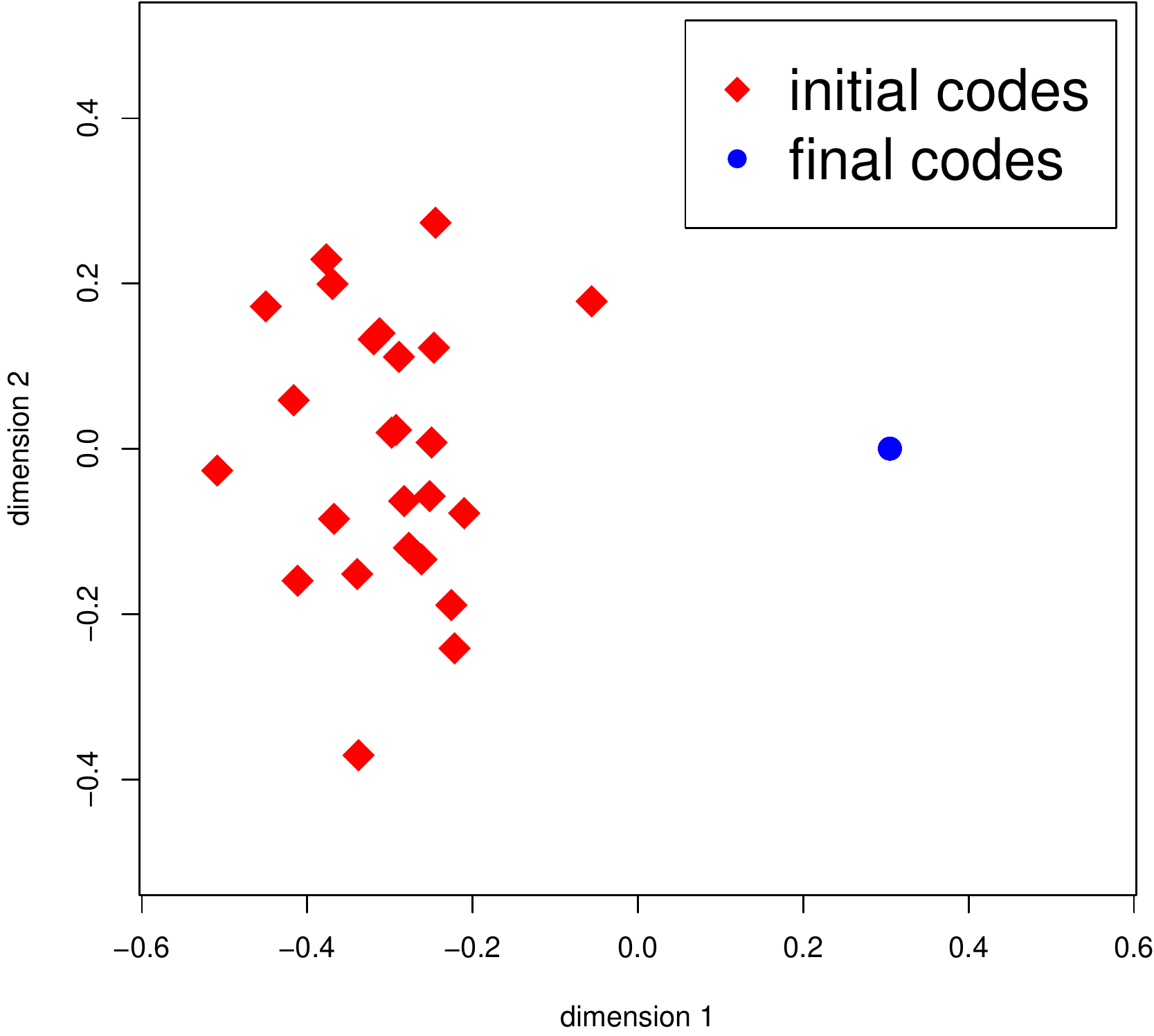}
	\caption{\doublespacing\small{2-dimensional plot of code distance: red points are codes at the beginning of the optimisation process; blue points
			are codes at the end of the optimisation process (where the distance between every pair of codes is zero).}}
  \label{fig:map_allvsall_25}
\end{figure}

In the first scenario, each agent $\theta_i$ perceives the output of every other possible agent $\theta_j$ with the same probability,
that is $p(\theta_i, \theta_j) = 1/25^2$ for every $i, j \in [1,25]$. The maximum average code similarity is bounded by
$I(Y_\Theta;Y_{\Theta^\prime}) = 1.71908$ bits, which is achieved under two conditions: first, every code must be a one-to-one mapping;
second, the code must be universal. This is indeed the outcome of the performed optimisation, as we show in Fig. \ref{fig:map_allvsall_25}:
the optimised codes (blue points) converged into a universal code (the distance between any of them is zero). Each red (diamond) point correspond
to an initial code.

\begin{figure}[htp]
	\centering
	\includegraphics[scale=0.35]{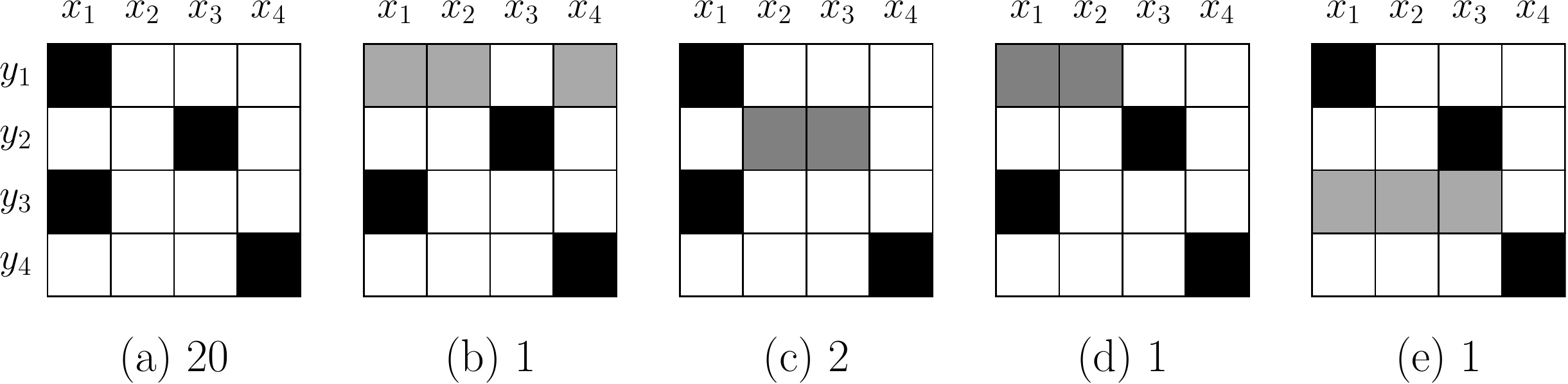}
	\caption{\doublespacing\small{Representation of the codes $p(x|y)$ by a heat-map using inverse grayscale. For each evolved code, we output the number of agents
			 adopting it. This code distribution was achieved with $25$ agents in a well-mixed population.}}
	\label{fig:allvsall_codes}
\end{figure}

The resulting code adopted by the population is a one-to-one mapping between sensor states and outputs, and any of the $24$ possible
one-to-one mappings is a global maximum (there are $4$ sensor states and $4$ possible outputs). However, it is still interesting to briefly analyse
the possible paths towards a universal and optimal code. In Fig. \ref{fig:allvsall_codes}, we show the distribution of the adopted codes by the
agents of the population in an iteration of the optimisation process where the average code similarity is $I(X_\Theta;X_{\Theta^\prime}) = 1.18276$
bits. Here, the most popular code is the suboptimal code shown in Fig. \ref{fig:allvsall_codes} (a). This results from the particular initialised
codes, driving the agents temporarily towards a suboptimal code. However, once any of the many-to-one codes becomes (nearly) universally adopted,
then any code's deviation improving the code similarity will eventually drive the convention towards optimality. The fact that it does not need
simultaneous changes in the code increases the likeliness of improving the code similarity.

\subsection{Spatially-structured population}
\label{sec:grid}

In another set-up, we assume the agents are structured in a $5\times5$ grid, where $p(\theta, \theta^\prime) = 1/105$ if $\theta$ and
$\theta^\prime$ are neighbours or when $\theta = \theta^\prime$ (see Fig. \ref{fig:gridcolours} for a representation of the structure).
After randomly initialising the codes, the performed optimisation plateaued on an average code similarity of $I(X_\Theta;X_{\Theta^\prime}) = 
1.13536$ bits. As in the former scenario, here the optimal solution is also a universal code with a one-to-one mapping. However, in this case,
the result is not a universal code, as can be appreciated in Fig. \ref{fig:map_grid_25}. Spatially structured populations are sensitive to the
initial codes and how codes are updated.

\definecolor{color1}{RGB}{27,158,119}
\definecolor{color2}{RGB}{217,95,2}
\definecolor{color3}{RGB}{117,112,179}
\definecolor{color4}{RGB}{231,41,138}
\definecolor{color5}{RGB}{102,166,30}
\definecolor{color6}{RGB}{230,171,2}
\definecolor{color7}{RGB}{166,118,29}
\definecolor{color8}{RGB}{102,102,102}

\begin{figure}[htp]
	\centering
	\begin{minipage}[b]{.49\textwidth}
		\centering
		\includegraphics[scale=0.4]{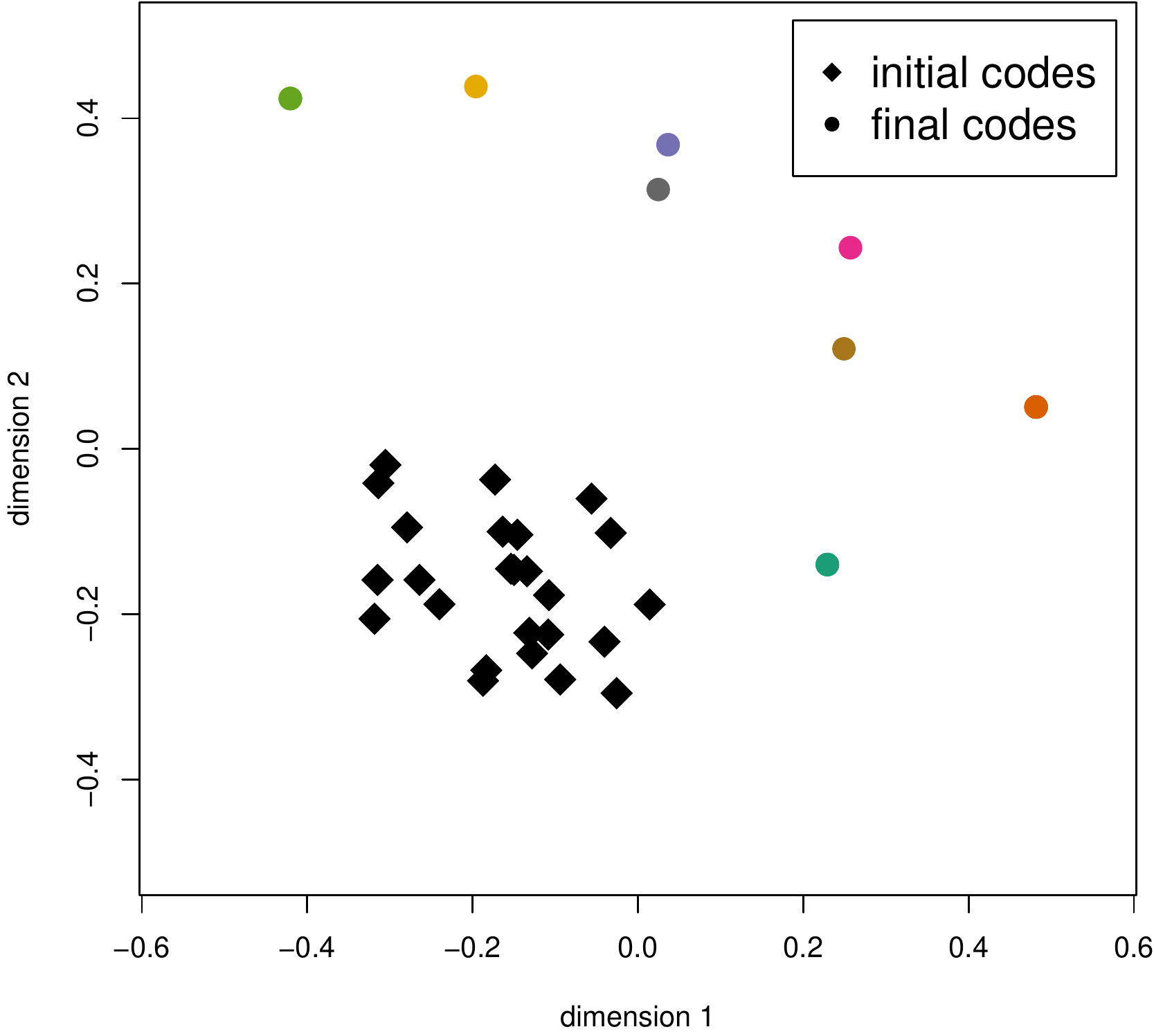}
		\caption{\doublespacing\small{2-dimensional plot of code distance: points in diamond shape represent codes at the beginning of the optimisation process; rounded points
				represent codes at the end of the optimisation process. The points are coloured in order to be able to relate this plot with the figure beside it.}}
		\label{fig:map_grid_25}
	\end{minipage}\hfill
	\begin{minipage}[b]{.49\textwidth}
		\centering
		  \begin{tikzpicture}
		    [->,>=stealth',shorten >=1pt,auto,node distance=1.2cm,
		    thick,main node/.style={font=\ttfamily\large\mdseries}]
		
			    \node[main node,fill=color1] (1) [] {a};
			    \node[main node,fill=color1] (2) [right of=1] {a};
			    \node[main node,fill=color4] (3) [right of=2] {b};
			    \node[main node,fill=color4] (4) [right of=3] {b};
			    \node[main node,fill=color4] (5) [right of=4] {b};
		
				\node[main node,fill=color1] (6) [below of=1] {a};
			    \node[main node,fill=color1] (7) [right of=6] {a};
				\node[main node,fill=color4] (8) [right of=7] {b};
				\node[main node,fill=color4] (9) [right of=8] {b};
			    \node[main node,fill=color4] (10) [right of=9] {b};
		
				\node[main node,fill=color3] (11) [below of=6] {c};
			    \node[main node,fill=color3] (12) [right of=11] {c};
				\node[main node,fill=color2] (13) [right of=12] {d};
				\node[main node,fill=color4] (14) [right of=13] {b};
			    \node[main node,fill=color4] (15) [right of=14] {b};
		
				\node[main node,fill=color5] (16) [below of=11] {e};
			    \node[main node,fill=color5] (17) [right of=16] {e};
				\node[main node,fill=color6] (18) [right of=17] {f};
				\node[main node,fill=color7] (19) [right of=18] {g};
			    \node[main node,fill=color4] (20) [right of=19] {b};
		
				\node[main node,fill=color5] (21) [below of=16] {e};
			    \node[main node,fill=color5] (22) [right of=21] {e};
				\node[main node,fill=color6] (23) [right of=22] {f};
				\node[main node,fill=color8] (24) [right of=23] {h};
			    \node[main node,fill=color2] (25) [right of=24] {d};
		
				\path[every node/.style={font=\sffamily\small}]
		
			      (1) edge node {} (2)
			      (2) edge node {} (1)
			      (2) edge node {} (3)
			      (3) edge node {} (2)
			      (3) edge node {} (4)
			      (4) edge node {} (3)
			      (4) edge node {} (5)
			      (5) edge node {} (4)
		
			      (6) edge node {} (7)
			      (7) edge node {} (6)
			      (7) edge node {} (8)
			      (8) edge node {} (7)
			      (8) edge node {} (9)
			      (9) edge node {} (8)
			      (9) edge node {} (10)
			      (10) edge node {} (9)
		
			      (11) edge node {} (12)
			      (12) edge node {} (11)
			      (12) edge node {} (13)
			      (13) edge node {} (12)
			      (13) edge node {} (14)
			      (14) edge node {} (13)
			      (14) edge node {} (15)
			      (15) edge node {} (14)
		
			      (16) edge node {} (17)
			      (17) edge node {} (16)
			      (17) edge node {} (18)
			      (18) edge node {} (17)
			      (18) edge node {} (19)
			      (19) edge node {} (18)
			      (19) edge node {} (20)
			      (20) edge node {} (19)
		
			      (21) edge node {} (22)
			      (22) edge node {} (21)
			      (22) edge node {} (23)
			      (23) edge node {} (22)
			      (23) edge node {} (24)
			      (24) edge node {} (23)
			      (24) edge node {} (25)
			      (25) edge node {} (24)
		
		
			      (1) edge node {} (6)
			      (6) edge node {} (1)
			      (2) edge node {} (7)
			      (7) edge node {} (2)
			      (3) edge node {} (8)
			      (8) edge node {} (3)
			      (4) edge node {} (9)
			      (9) edge node {} (4)
			      (5) edge node {} (10)
			      (10) edge node {} (5)
		
			      (11) edge node {} (6)
			      (6) edge node {} (11)
			      (12) edge node {} (7)
			      (7) edge node {} (12)
			      (13) edge node {} (8)
			      (8) edge node {} (13)
			      (14) edge node {} (9)
			      (9) edge node {} (14)
			      (15) edge node {} (10)
			      (10) edge node {} (15)
		
			      (11) edge node {} (16)
			      (16) edge node {} (11)
			      (12) edge node {} (17)
			      (17) edge node {} (12)
			      (13) edge node {} (18)
			      (18) edge node {} (13)
			      (14) edge node {} (19)
			      (19) edge node {} (14)
			      (15) edge node {} (20)
			      (20) edge node {} (15)
		
			      (21) edge node {} (16)
			      (16) edge node {} (21)
			      (22) edge node {} (17)
			      (17) edge node {} (22)
			      (23) edge node {} (18)
			      (18) edge node {} (23)
			      (24) edge node {} (19)
			      (19) edge node {} (24)
			      (25) edge node {} (20)
			      (20) edge node {} (25)
		
			      (1) edge [loop left] node {} (1)
			      (6) edge [loop left] node {} (6)
			      (11) edge [loop left] node {} (11)
			      (16) edge [loop left] node {} (16)
			      (21) edge [loop left] node {} (21)
		
			      (2) edge [loop above] node {} (2)
			      (3) edge [loop above] node {} (3)
			      (4) edge [loop above] node {} (4)
		
			      (5) edge [loop right] node {} (5)
			      (10) edge [loop right] node {} (10)
			      (15) edge [loop right] node {} (15)
			      (20) edge [loop right] node {} (20)
			      (25) edge [loop right] node {} (25)
		
			      (22) edge [loop below] node {} (22)
			      (23) edge [loop below] node {} (23)
			      (24) edge [loop below] node {} (24)
		
			      (7) edge [in=150,out=120,loop] node {} (7)
			      (12) edge [in=150,out=120,loop] node {} (12)
			      (17) edge [in=150,out=120,loop] node {} (17)
		
			      (8) edge [in=60,out=30,loop] node {} (8)
			      (9) edge [in=60,out=30,loop] node {} (9)
			      (13) edge [in=60,out=30,loop] node {} (13)
			      (14) edge [in=60,out=30,loop] node {} (14)
			      (18) edge [in=60,out=30,loop] node {} (18)
			      (19) edge [in=60,out=30,loop] node {} (19)
		
				  ;
		
		  \end{tikzpicture}
		  \caption{\doublespacing\small{Representation of the spatial structure utilised for the experiment. Agents are assumed to be distributed in a grid:
		          an edge from one agent to another means that one agent perceives the output of the other. Agents are labelled (see Fig. \ref{fig:grid_codes})
				  and coloured according to their adopted code.}}
		  \label{fig:gridcolours}
	\end{minipage}
\end{figure}

\begin{figure}[htp]
	\centering
	\includegraphics[scale=0.35]{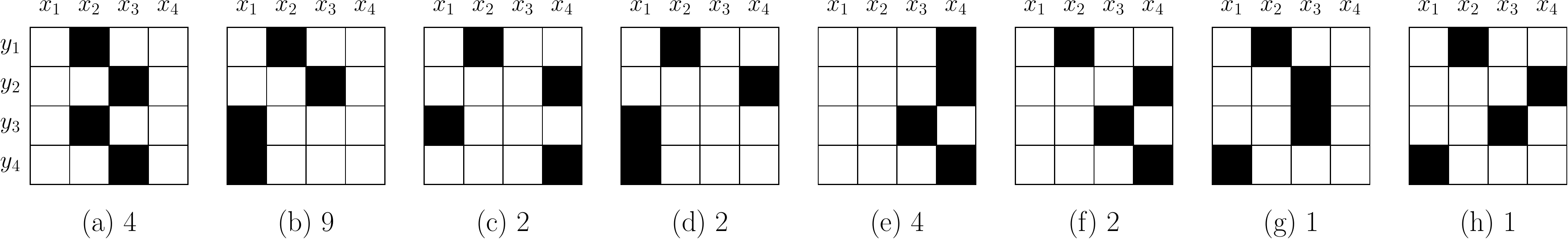}
	\caption{\doublespacing\small{Representation of the codes $p(x|y)$ by a heat-map using inverse grey scale. For each evolved code, we output the number of agents
			 adopting it. This code distribution was achieved with $25$ agents in a grid structure.}}
	\label{fig:grid_codes}
\end{figure}

The resulting code distribution among the population is shown in Fig. \ref{fig:grid_codes}, with $8$ different codes in the population.
Where well-mixed populations evolved the use of common codes, agreement on codes only occurred among neighbours in spatially structured
populations. As a consequence, many local conventions are established within neighbourhoods, and, once this situation is reached, the improvement
of the total code similarity requires simultaneous changes to the agent's codes. For instance, the code shown in Fig. \ref{fig:grid_codes} (e)
could increase the average similarity of the population if $p(x_2|y_1) = 1$, as it is in the rest of the codes. However, for this to happen
(in this particular case), at least two agents need to change their code simultaneously (otherwise the average similarity decreases), which
makes the deviation from the resulting code distribution unlikely.

\subsection{Flexible population structure}
\label{sec:free}

\begin{figure}[htp]
	\centering
	\includegraphics[scale=0.65]{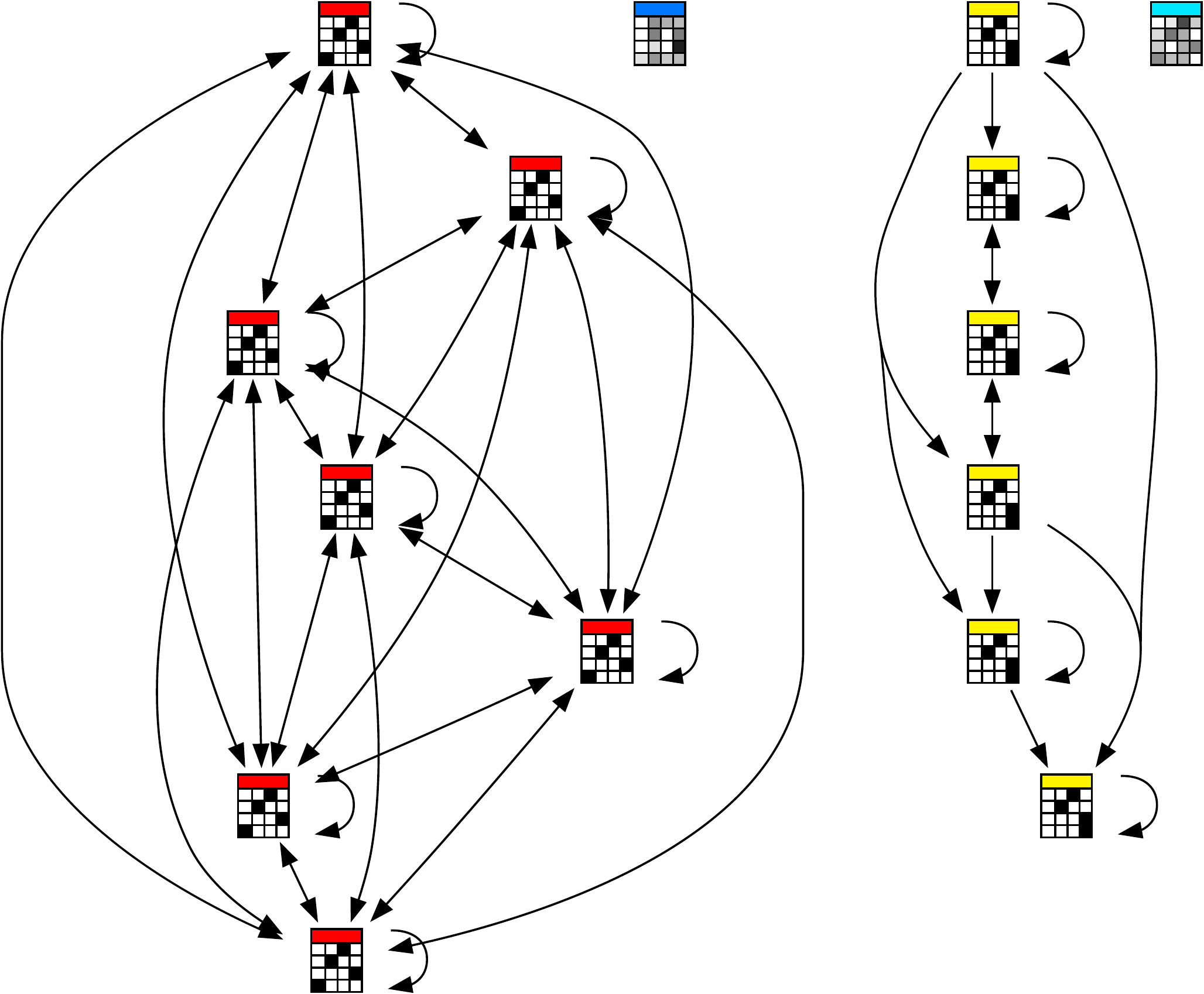}
	\caption{\doublespacing\small{Each node in the graph corresponds to the code of an agent. There is a weighted edge between agent $\theta_i$ and $\theta_j$
			if $p(\theta_i, \theta_j) > 0$ (which is the weight). We omit weights of edges in the graph since they all are roughly of similar value.
			The temperature colours on top of the nodes indicate the amount of environmental information
			they would contribute to any agent perceiving only that agents output.}}
  \label{fig:free_structure_codes}
\end{figure}

For the third scenario, we let the structure co-evolve with the codes without any constraint (the probability distribution of the interaction
between agents, $p(\Xi)$, is optimised together with the codes). In this case, the resulting average code similarity
is nearly optimal, but the code is not necessarily universal. This is because, when the structure is not fixed, agents form roughly disconnected
clusters of related codes. In this process, the interaction probability of agents with unrelated codes will vanish. However, once the clusters are
formed, if it is not a single isolated agent (such that no other agent perceives its output), then codes of agents are universal within each cluster.
This is exemplified by the code distribution and population structure we obtained (see Fig. \ref{fig:free_structure_codes}). Here, we have two
clusters with universal codes, one optimal (in red) and the other suboptimal (in yellow). Agents with dissimilar codes from every other agent
they interact with will become isolated in the optimisation process, as the example shows for two agents (light and dark blue). 

To summarise, the optimal code similarity equals $I(Y_\Theta;Y_{\Theta^\prime})$, and is achieved, for instance, when all agents adopt the same
one-to-one mapping. Nevertheless, the interaction probability allows agents to form disconnected clusters of related codes, where several
one-to-one mappings could result while still achieving optimality. Theoretically, we could have as many one-to-one mappings as the minimum
between the amount of agents and the total one-to-one mapping combinations ($24$ in this case).

\subsection{Emerging concepts in a well-mixed heterogeneous population}
\label{sec:wmhetpop}

So far, we have only considered populations of agents that acquired the same aspects of information from $\mu$ (\emph{i.e.}, 
$p(Y_{\theta_i}|\mu) = p(Y_{\theta_j}|\mu)$ for any pair of agents $\theta_i$, $\theta_j$). The assumption was that the information that was relevant
for the survival of the agents was the same among the agents of the population, and this was represented by $\mu$. Now, we consider a more general
scenario, where different types of agents acquire different aspects from the environmental conditions $\mu$. We investigate whether it is possible
for an agent that does not directly perceive the environment at all (we call this type of agent ``blind'') to predict conditions based solely on the
outputs of other agents. We consider a well-mixed population, such that different types of agents are forced to talk to each other. Considerations
with a flexible population structure are not interesting for our purposes, since in these cases, each type of agent forms a cluster disconnected
from clusters of other types. This was confirmed by simulations which are not shown here.

Let us illustrate the idea with a relatively simple scenario: we consider five types of agents (we denote the i-\emph{th} type $\phi_i$), where each
type can only distinguish whether the current state of the environment belongs to its coloured region or not. The environment consists of $9$ states,
and the probability of each state is uniformly distributed. We illustrate this environment by a $3 \times 3$ grid, as shown in
Fig. \ref{fig:sensor_types}, although the square does not denote the physical structure of the environment. Then, the outputs of each type of agent will
be related to the regions they capture. For instance, for agents of type $\phi_2$ with the same deterministic code, if
$Pr(\mu \in \{ 1, 2, 4, 5 \} | X_\theta = x)$ equals one (for all $\theta$ of type $\phi_2$), then $x$ will signify that this agent is currently in
the region coloured in red in Fig. \ref{fig:sensor_types}. We say that a population of agents has a joint \emph{concept} of the environment if by
considering its representation of the environmental information they capture, we can obtain information about the environment, \emph{i.e.} we
require that $I(\mu;X_\Theta) > 0$. For instance, the symbol $x$ in the example above, assuming that it is only utilised by agents of the same
type, can be understood as representing the concept ``top-left'' of the grid. 

\definecolor{color22}{RGB}{228,26,28}
\definecolor{color23}{RGB}{77,175,74}
\definecolor{color24}{RGB}{55,126,184}
\definecolor{color25}{RGB}{152,78,163}

\begin{figure}[ht]
	\centering
	\begin{tikzpicture}
		\def\bc{1.5} 
		\def\as{1} 
		\def\code{0}
		\filldraw[draw=black,thick,fill={rgb,1:red,1;green,1;blue,1}] (0+\code+\code*\bc,-0) rectangle (0.5+\code+\code*\bc,-0.5);
		\filldraw[draw=black,thick,fill={rgb,1:red,1;green,1;blue,1}] (0.5+\code+\code*\bc,-0) rectangle (1.0+\code+\code*\bc,-0.5);
		\filldraw[draw=black,thick,fill={rgb,1:red,1;green,1;blue,1}] (1.0+\code+\code*\bc,-0) rectangle (1.5+\code+\code*\bc,-0.5);
		\filldraw[draw=black,thick,fill={rgb,1:red,1;green,1;blue,1}] (0+\code+\code*\bc,-0.5) rectangle (0.5+\code+\code*\bc,-1.0);
		\filldraw[draw=black,thick,fill={rgb,1:red,1;green,1;blue,1}] (0.5+\code+\code*\bc,-0.5) rectangle (1.0+\code+\code*\bc,-1.0);
		\filldraw[draw=black,thick,fill={rgb,1:red,1;green,1;blue,1}] (1.0+\code+\code*\bc,-0.5) rectangle (1.5+\code+\code*\bc,-1.0);
		\filldraw[draw=black,thick,fill={rgb,1:red,1;green,1;blue,1}] (0+\code+\code*\bc,-1.0) rectangle (0.5+\code+\code*\bc,-1.5);
		\filldraw[draw=black,thick,fill={rgb,1:red,1;green,1;blue,1}] (0.5+\code+\code*\bc,-1.0) rectangle (1.0+\code+\code*\bc,-1.5);
		\filldraw[draw=black,thick,fill={rgb,1:red,1;green,1;blue,1}] (1.0+\code+\code*\bc,-1.0) rectangle (1.5+\code+\code*\bc,-1.5);
		\node[] at (0.25+\code+\code*\bc, -0.25) {$1$};
		\node[] at (0.75+\code+\code*\bc, -0.25) {$2$};
		\node[] at (1.25+\code+\code*\bc, -0.25) {$3$};
		\node[] at (0.25+\code+\code*\bc, -0.75) {$4$};
		\node[] at (0.75+\code+\code*\bc, -0.75) {$5$};
		\node[] at (1.25+\code+\code*\bc, -0.75) {$6$};
		\node[] at (0.25+\code+\code*\bc, -1.25) {$7$};
		\node[] at (0.75+\code+\code*\bc, -1.25) {$8$};
		\node[] at (1.25+\code+\code*\bc, -1.25) {$9$};
		\node[] at (0.75+\code+\code*\bc, -2.) {states of $\mu$};
		\def\code{1}
		\filldraw[draw=black,thick,fill={rgb,1:red,0.33;green,0.33;blue,0.33}] (0+\code+\code*\bc,-0) rectangle (0.5+\code+\code*\bc,-0.5);
		\filldraw[draw=black,thick,fill={rgb,1:red,0.33;green,0.33;blue,0.33}] (0.5+\code+\code*\bc,-0) rectangle (1.0+\code+\code*\bc,-0.5);
		\filldraw[draw=black,thick,fill={rgb,1:red,0.33;green,0.33;blue,0.33}] (1.0+\code+\code*\bc,-0) rectangle (1.5+\code+\code*\bc,-0.5);
		\filldraw[draw=black,thick,fill={rgb,1:red,0.33;green,0.33;blue,0.33}] (0+\code+\code*\bc,-0.5) rectangle (0.5+\code+\code*\bc,-1.0);
		\filldraw[draw=black,thick,fill={rgb,1:red,0.33;green,0.33;blue,0.33}] (0.5+\code+\code*\bc,-0.5) rectangle (1.0+\code+\code*\bc,-1.0);
		\filldraw[draw=black,thick,fill={rgb,1:red,0.33;green,0.33;blue,0.33}] (1.0+\code+\code*\bc,-0.5) rectangle (1.5+\code+\code*\bc,-1.0);
		\filldraw[draw=black,thick,fill={rgb,1:red,0.33;green,0.33;blue,0.33}] (0+\code+\code*\bc,-1.0) rectangle (0.5+\code+\code*\bc,-1.5);
		\filldraw[draw=black,thick,fill={rgb,1:red,0.33;green,0.33;blue,0.33}] (0.5+\code+\code*\bc,-1.0) rectangle (1.0+\code+\code*\bc,-1.5);
		\filldraw[draw=black,thick,fill={rgb,1:red,0.33;green,0.33;blue,0.33}] (1.0+\code+\code*\bc,-1.0) rectangle (1.5+\code+\code*\bc,-1.5);
		\node[] at (0.75+\code+\code*\bc, -2.) {type $\phi_1$};
		\def\code{2}
		\filldraw[draw=black,thick,fill=color22] (0+\code+\code*\bc,-0) rectangle (0.5+\code+\code*\bc,-0.5);
		\filldraw[draw=black,thick,fill=color22] (0.5+\code+\code*\bc,-0) rectangle (1.0+\code+\code*\bc,-0.5);
		\filldraw[draw=black,thick,fill={rgb,1:red,1;green,1;blue,1}] (1.0+\code+\code*\bc,-0) rectangle (1.5+\code+\code*\bc,-0.5);
		\filldraw[draw=black,thick,fill=color22] (0+\code+\code*\bc,-0.5) rectangle (0.5+\code+\code*\bc,-1.0);
		\filldraw[draw=black,thick,fill=color22] (0.5+\code+\code*\bc,-0.5) rectangle (1.0+\code+\code*\bc,-1.0);
		\filldraw[draw=black,thick,fill={rgb,1:red,1;green,1;blue,1}] (1.0+\code+\code*\bc,-0.5) rectangle (1.5+\code+\code*\bc,-1.0);
		\filldraw[draw=black,thick,fill={rgb,1:red,1;green,1;blue,1}] (0+\code+\code*\bc,-1.0) rectangle (0.5+\code+\code*\bc,-1.5);
		\filldraw[draw=black,thick,fill={rgb,1:red,1;green,1;blue,1}] (0.5+\code+\code*\bc,-1.0) rectangle (1.0+\code+\code*\bc,-1.5);
		\filldraw[draw=black,thick,fill={rgb,1:red,1;green,1;blue,1}] (1.0+\code+\code*\bc,-1.0) rectangle (1.5+\code+\code*\bc,-1.5);
		\node[] at (0.75+\code+\code*\bc, -2.) {type $\phi_2$};
		\def\code{3}
		\filldraw[draw=black,thick,fill={rgb,1:red,1;green,1;blue,1}] (0+\code+\code*\bc,-0) rectangle (0.5+\code+\code*\bc,-0.5);
		\filldraw[draw=black,thick,fill=color23] (0.5+\code+\code*\bc,-0) rectangle (1.0+\code+\code*\bc,-0.5);
		\filldraw[draw=black,thick,fill=color23] (1.0+\code+\code*\bc,-0) rectangle (1.5+\code+\code*\bc,-0.5);
		\filldraw[draw=black,thick,fill={rgb,1:red,1;green,1;blue,1}] (0+\code+\code*\bc,-0.5) rectangle (0.5+\code+\code*\bc,-1.0);
		\filldraw[draw=black,thick,fill=color23] (0.5+\code+\code*\bc,-0.5) rectangle (1.0+\code+\code*\bc,-1.0);
		\filldraw[draw=black,thick,fill=color23] (1.0+\code+\code*\bc,-0.5) rectangle (1.5+\code+\code*\bc,-1.0);
		\filldraw[draw=black,thick,fill={rgb,1:red,1;green,1;blue,1}] (0+\code+\code*\bc,-1.0) rectangle (0.5+\code+\code*\bc,-1.5);
		\filldraw[draw=black,thick,fill={rgb,1:red,1;green,1;blue,1}] (0.5+\code+\code*\bc,-1.0) rectangle (1.0+\code+\code*\bc,-1.5);
		\filldraw[draw=black,thick,fill={rgb,1:red,1;green,1;blue,1}] (1.0+\code+\code*\bc,-1.0) rectangle (1.5+\code+\code*\bc,-1.5);
		\node[] at (0.75+\code+\code*\bc, -2.) {type $\phi_3$};
		\def\code{4}
		\filldraw[draw=black,thick,fill={rgb,1:red,1;green,1;blue,1}] (0+\code+\code*\bc,-0) rectangle (0.5+\code+\code*\bc,-0.5);
		\filldraw[draw=black,thick,fill={rgb,1:red,1;green,1;blue,1}] (0.5+\code+\code*\bc,-0) rectangle (1.0+\code+\code*\bc,-0.5);
		\filldraw[draw=black,thick,fill={rgb,1:red,1;green,1;blue,1}] (1.0+\code+\code*\bc,-0) rectangle (1.5+\code+\code*\bc,-0.5);
		\filldraw[draw=black,thick,fill=color24] (0+\code+\code*\bc,-0.5) rectangle (0.5+\code+\code*\bc,-1.0);
		\filldraw[draw=black,thick,fill=color24] (0.5+\code+\code*\bc,-0.5) rectangle (1.0+\code+\code*\bc,-1.0);
		\filldraw[draw=black,thick,fill={rgb,1:red,1;green,1;blue,1}] (1.0+\code+\code*\bc,-0.5) rectangle (1.5+\code+\code*\bc,-1.0);
		\filldraw[draw=black,thick,fill=color24] (0+\code+\code*\bc,-1.0) rectangle (0.5+\code+\code*\bc,-1.5);
		\filldraw[draw=black,thick,fill=color24] (0.5+\code+\code*\bc,-1.0) rectangle (1.0+\code+\code*\bc,-1.5);
		\filldraw[draw=black,thick,fill={rgb,1:red,1;green,1;blue,1}] (1.0+\code+\code*\bc,-1.0) rectangle (1.5+\code+\code*\bc,-1.5);
		\node[] at (0.75+\code+\code*\bc, -2.) {type $\phi_4$};
		\def\code{5}
		\filldraw[draw=black,thick,fill={rgb,1:red,1;green,1;blue,1}] (0+\code+\code*\bc,-0) rectangle (0.5+\code+\code*\bc,-0.5);
		\filldraw[draw=black,thick,fill={rgb,1:red,1;green,1;blue,1}] (0.5+\code+\code*\bc,-0) rectangle (1.0+\code+\code*\bc,-0.5);
		\filldraw[draw=black,thick,fill={rgb,1:red,1;green,1;blue,1}] (1.0+\code+\code*\bc,-0) rectangle (1.5+\code+\code*\bc,-0.5);
		\filldraw[draw=black,thick,fill={rgb,1:red,1;green,1;blue,1}] (0+\code+\code*\bc,-0.5) rectangle (0.5+\code+\code*\bc,-1.0);
		\filldraw[draw=black,thick,fill=color25] (0.5+\code+\code*\bc,-0.5) rectangle (1.0+\code+\code*\bc,-1.0);
		\filldraw[draw=black,thick,fill=color25] (1.0+\code+\code*\bc,-0.5) rectangle (1.5+\code+\code*\bc,-1.0);
		\filldraw[draw=black,thick,fill={rgb,1:red,1;green,1;blue,1}] (0+\code+\code*\bc,-1.0) rectangle (0.5+\code+\code*\bc,-1.5);
		\filldraw[draw=black,thick,fill=color25] (0.5+\code+\code*\bc,-1.0) rectangle (1.0+\code+\code*\bc,-1.5);
		\filldraw[draw=black,thick,fill=color25] (1.0+\code+\code*\bc,-1.0) rectangle (1.5+\code+\code*\bc,-1.5);
		\node[] at (0.75+\code+\code*\bc, -2.) {type $\phi_5$};
	\end{tikzpicture}
	\caption{\doublespacing\small{Representation of the conditional probabilities $p(Y_\theta|\mu)$ for an agent $\theta$ of each type. These are defined such that
             each type of agent can only distinguish between the coloured region and the white region. For instance, the sensor of type $\phi_2$ is defined
             as $Pr(Y = y_1 | \mu) = 1$ if $\mu \in \{ 1, 2, 4, 5 \}$, and zero otherwise, and $Pr(Y = y_2 | \mu) = 1$ if $\mu \notin \{1, 2, 4, 5 \}$,
             and zero otherwise. For type $\phi_1$, $Pr(Y = y_1 | \mu) = 0.5$ and $Pr(Y = y_2 | \mu) = 0.5$ ($|Y| = 2$ for all types of agents).}}
  \label{fig:sensor_types}
\end{figure}
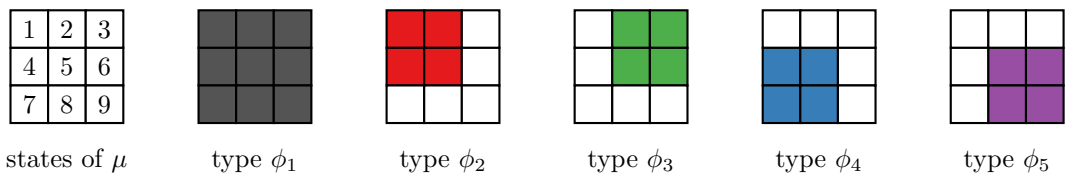

The amount of environmental information that an agent $\theta$ of type $\phi_1$ (a blind agent) captures is $I(\mu;Y_{\theta}) = 0$ bits, while all
agents $\theta$ of the other types capture $I(\mu;Y_{\theta}) = 0.991076$ bits (note that the total entropy in $\mu$ to be resolved is
$H(\mu) = 3.16993$ bits). Throughout this study, we considered that agents predict the environment by considering their perceptions together with the outputs
of other agents. The blind agent, instead, since it is not able to capture any direct cue from $\mu$, we consider capable of perceiving the outputs of
both of the agents selected by $\Theta$ and $\Theta^\prime$. With this relaxed consideration, we say a blind agent \emph{has a concept} of the
environment if $I(\mu;X_\Theta,X_{\Theta^\prime}) > 0$, \emph{i.e.} we consider the maximum amount of information an agent can possibly infer from the
joint outputs $X_\Theta$ and $X_{\Theta^\prime}$. 

Let us recall that the structure of the population is well-mixed, and thus the distribution of outputs of all agents is considered, including the
blind ones, which are not able to express (via their outputs) any particular concept by themselves (for a blind agent $\theta$, $I(\mu;X_\theta)
\leq I(\mu;Y_\theta) = 0$, \emph{i.e.} $I(\mu;X_\Theta)$ vanishes). Therefore, whether a blind agent has some concept of the environment will depend,
first, on the universality of the codes of each type of agent (agents representing the same information with different symbols may create ambiguities).
Second, on the cardinality of the alphabet of $X$ (\emph{i.e.} $|X|$) utilised by the population. A small alphabet will force agents to represent
different concepts of the environment with the same symbols, while a large alphabet is likely to result in exclusive representations of concepts for
each type of agent.

Taking this into account, we ask, is it possible for a blind agent to identify concepts of the environment? If so, how are these concepts related to
the concepts of the individual agents (other than the blind ones)? Is the size of the available alphabet related to the quality of the concepts?

\begin{figure}[htp]
	\centering
	\begin{minipage}[b]{.58\textwidth}
		\centering
		\renewcommand{\arraystretch}{1.2}
		\begin{tabular}{ c  c  c }
			$|X|$ & $I(\mu;X_\Theta,X_{\Theta^\prime})$ \\
			\hline
			$2$ & $0.34621$ \\
			$3$ & $0.56555$ \\
			$4$ & $0.71620$ \\
			$5$ & $0.95467$ \\
			$6$ & $1.08139$ \\
			$7$ & $1.18362$ \\
			$8$ & $1.30919$ \\
			$9$ & $1.30919$ \\
		\end{tabular}
		\captionof{table}{\doublespacing\small{Results of experiments where the size of the alphabet of a population varies. The maximum amount of environmental information
                       that a blind agent can infer is achieved with $|X| = 8$ and remains equal for bigger alphabets. As the size of the alphabet
                       decreases, this information also decreases.}}
		\label{tab:mi}
	\end{minipage}\hfill
	\begin{minipage}[b]{.38\textwidth}
		\centering
		\includegraphics[scale=0.25]{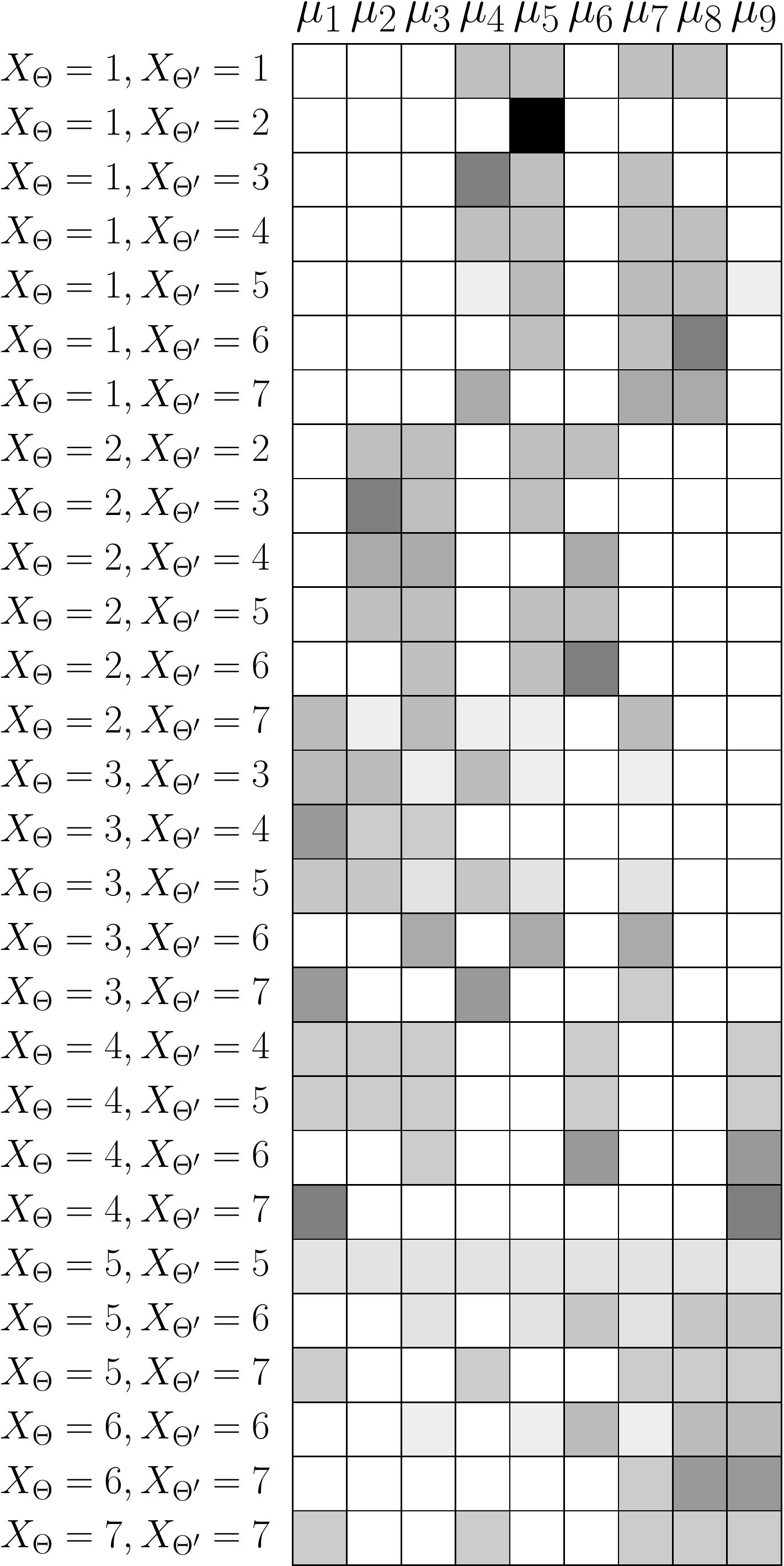}
		\captionof{figure}{\doublespacing\small{Conditional probability $p(\mu|X_\Theta,X_{\Theta^\prime})$ in inverse grey-scale. Each row represents a combination of values of $X_\Theta$ and
                 $X_{\Theta^\prime}$, and each column represents a state of $\mu$.}}
		\label{fig:concepts}
	\end{minipage}
\end{figure}

To study these questions, we performed different experiments varying the size of the alphabet $|X|$, where the rest of the parameters remained the same.
In these experiments, we optimised the similarity of codes for a population composed of $20$ agents, with $4$ agents of each of the five types. 
In Table \ref{tab:mi} we show that the cardinality of the alphabet of $X$ affects the limit of the amount of information a blind agent can possibly infer
about the environment.

Now, if we measure the uncertainty of the environment for a blind agent for each combination of outputs $X_\Theta$ and $X_{\Theta^\prime}$, we find that
for some of them, it is zero. For instance, with $|X| = 7$, we found that when $Pr(\mu = 5 | X_\Theta = 1, X_{\Theta^\prime} = 2) = 1.0$
(see Fig. \ref{fig:concepts}, where only combinations with $X_\Theta \leq X_{\Theta^\prime}$ are shown). These distributions are also valid when swapping
the values of $X_\Theta$ and $X_{\Theta^\prime}$, since in the well-mixed population the structure is symmetric.
Looking at the example of the conditional probability in Fig. \ref{fig:concepts}, we can find many other concepts, although none of them ---apart from
the one already discussed--- can uniquely identify a state of the environment. For instance, we have that $Pr(\mu | X_\Theta = 3, X_{\Theta^\prime} = 6) = 0.33$
when $\mu \in \{ 3, 5, 7 \}$, which is a concept for being on a particular diagonal of the environment.

\begin{figure}[htp]
	\centering
	\includegraphics[scale=0.75]{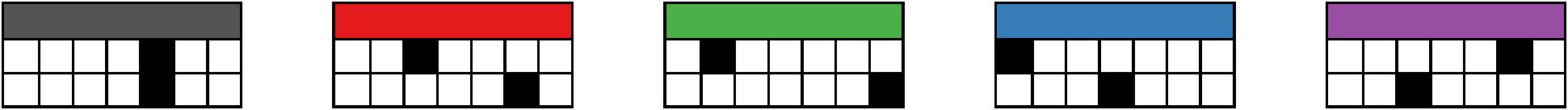}
	\caption{\doublespacing\small{Representation of codes $p(X_\Theta|Y_\Theta,\Theta)$ by a heat-map using inverse grayscale for the experiment with $|X| = 7$. For each node, the rows
            represent a sensor state $y$, while the columns represent an output state $x$. The colours on top of the nodes are used to distinguish the type of agent to whom
            the code belongs, and colours are related to those shown in Fig. \ref{fig:sensor_types}.}}
  \label{fig:cluster_concepts}
\end{figure}

In Fig. \ref{fig:cluster_concepts} we show the resulting codes (which are universal for each type, including the blind one) for this particular experiment.
Here, the types $\phi_2$ (red) and $\phi_5$ (purple) utilise the same symbols to represent different environmental conditions. By using a small size of the
alphabet for $X$, we force ambiguities in the population, but these will be chosen (by evolution) such that they are minimal. In this way, we maximise the
amount of information we can infer from the outputs (although this can be a local optimum). For instance, the outputs of the blind agents (type $\phi_1$) for
all the experiments never overlapped that of other types (unless we use $|X| = 2$, where there is no choice). In other words, blind agents always choose one
symbol so that they minimise the amount of utilised symbols from the whole population.

In all the performed experiments, we found that for values of $|X| \geq 6$, the blind agent can perfectly predict the environmental state $\mu = 5$ for
at least one combination of outputs $X_\Theta$ and $X_{\Theta^\prime}$. Interestingly, this new concept, which in this particular experiment can be called
the ``centre'' of the world or environment, cannot be obtained by looking to individual concepts only.

\section{Discussion}
\label{sec:discussion}

We considered four different scenarios of code evolution: in the first one, all agents perceived the outputs of all other agents,
including itself. We argued that two main stages of evolution can be recognised: in the first stage, a universal code is established,
which can be optimal or not. If it is not optimal, then a second stage will achieve optimality. The same result was obtained in 
\cite{Vetsigian2006}, in a model of the evolution of the genetic code (represented as a probabilistic mapping between codons and amino
acids), although universality and optimality were simultaneously achieved.

In the mentioned work, which developed further the ideas of \cite{Woese2002, Woese2004}, the authors argue that the universality of
the genetic code is a consequence of early communal evolution, mediated by horizontal gene transfer (HGT) between primitive cells. In
this evolutionary process, they argue, larger communities will have access (through the exchange of genetic material) to more innovations,
leading to faster evolution than smaller ones. Then, ``it is not better genetic codes that give an advantage but more common ones''
\cite{Vetsigian2006}. Although their model does not explicitly show this property, it is captured in our model. We show that a
more common, but not optimal code is widely adopted within a population (see Fig. \ref{fig:allvsall_codes}). However, in our model,
a code imposes itself as universal not because it provides access to more innovations (in our model there is no ``code exchange'', only
the outputs are shared), but because the population structure forces the adoption of the most popular code. After this stage, further
changes in the code of the agents eventually lead to optimality. 

In another related work, \cite{Oudeyer2005} explored the origins of language in a scenario consisting of artificial agents with a coupled
perception and production of speech sounds. Although this work is focused on plausible mechanisms for the origin of language, it assumes
the same similarity principle as we do (hearing a vocalisation increases the probability of producing similar vocalisations), arriving to
the same outcome (a universal language, or code). Other works have considered similar principles in the evolution of languages: for instance,
the naming game \cite{Steels1995} and the imitation game \cite{Boer2000}. However, these models assume some common conventions in order
to evolve new ones. In this study, our main assumption was that the population of agents depended on common environmental conditions.

Our second scenario, where the structure of the population is a grid, showed how establishing local conventions in early stages of evolution
constrains the outcome of the code distribution, since to reconcile different conventions, several simultaneous changes are needed. On the
other hand, in our third scenario, where we let the structure of the population change simultaneously with the codes themselves, such
situations are avoided by ``disconnecting'' clusters with dissimilar conventions. This property enhances evolution, and can potentially lead
to the adoption of several different conventions within an increasingly fragmenting, or ``speciating'' population. 

Our last scenario assumed perceptual constraints on the environmental information of each agent, an we looked at emerging concepts within a
well-mixed population. This scenario was studied in \cite{Moller2008}, where, as well as in our study, new conceptualisations of the world
emerged as a result of considering together the concepts of every agent. In both studies, the new concept was not representable individually
by any agent. Differently from the mentioned study, the new concepts obtained in our study were the result of a simple similarity maximisation
principle, while in the work of \cite{Moller2008}, concepts were obtained through the modelling of an explicit fitness function. 

The evolution of conventional codes could be interpreted, in the widest sense, as a form of cultural evolution. For instance, considering the
definition of culture given by \cite{Richerson2005}: \emph{``Culture is information capable of affecting individuals' behavior that they
acquire from other members of their species through teaching, imitation, and other forms of social transmission.''}, it could be argued that
a form of cultural information is present in organisms, such as bacteria or plants. Although there is a dependence among the different dimensions
on which information is transmitted in organisms (if we assume the dimensions to be, for instance, genetic, epigenetic, behavioural and
symbol-based, as proposed by \cite{Jablonka2005}), our model assumes freedom of choice in one dimension, without direct influence on the
others.

Finally, communication between individuals of a population opens up the possibility of ``signal cheaters'', which could be either individuals
that do not produce signals themselves but still perceive those of the others, or individuals who exploit other individual's learned responses
to symbols to their advantage. However, our model does not allow such behaviour, since the code producing the outputs functions, implicitly,
as the interpreter of the perceived signals.

\section{Conclusion}
\label{sec:conclusion}

In the proposed model, we introduced a key assumption which allowed us to evolve, for some structures, universal and optimal codes.
This assumption states that an agent cannot distinguish the sources of the outputs it perceives from other agents. Following from
this, a universal code will necessary introduce semantics by relating symbols to environmental conditions (via the internal states of the
agent). Our model proposes an information-theoretic way of measuring the similarity within a population of codes.

In this work, we proposed, as an evolutionary principle, that agents try to maximise their side information about the environment indirectly
by maximising their mutual code similarity. This behaviour produces several interesting outcomes in the code distribution of a structured
population. Depending on the population structure, it captures the evolution of a universal and optimal code (well-mixed population structure),
while also the evolution of different codes organised in clusters (in a freely evolving structure), which allows the establishment of optimal as
well as suboptimal conventions.

Finally, we considered a well-mixed heterogeneous population with perceptual constraints on the agents about the environment, and showed how,
just by looking at the outputs of agents, it is possible to extract concepts that relate to the environment, concepts that none of the agents
of the population could individually represent.

\bibliographystyle{plain}
\bibliography{burgos}

\end{document}